\documentclass[aps,prl,twocolumn,byrevtex,showpacs,preprintnumbers,amsmath,amssymb]{revtex4}

\usepackage{graphicx}
\usepackage{color}

\begin{document}
\title{Coherent reflection of He atom beams from rough surfaces at near-grazing incidence}

\author{Bum Suk Zhao}
\author{H. Christian Schewe}
\author{Gerard Meijer}
\author{Wieland Sch\"ollkopf}
\email{wschoell@fhi-berlin.mpg.de} \affiliation{Fritz-Haber-Institut
der Max-Planck-Gesellschaft, Faradayweg 4-6, 14195 Berlin, Germany}
\date{\today}

\begin{abstract} We here report coherent reflection of thermal He atom
beams from various microscopically rough surfaces at grazing
incidence. For a sufficiently small normal component $k_z$ of the
incident wave-vector of the atom the reflection probability is found
to be a function of $k_z$ only. This behavior is explained by
quantum-reflection at the attractive branch of the Casimir-van der
Waals interaction potential. For larger values of $k_z$ the overall
reflection probability decreases rapidly and is found to also depend
on the parallel component $k_x$ of the wave-vector. The material
specific $k_x$ dependence for this classical reflection at the
repulsive branch of the potential is explained qualitatively in
terms of the averaging-out of the surface roughness under grazing
incidence conditions.

\end{abstract}


\maketitle

Coherent reflection of an atom from a solid surface can happen via
two different mechanisms; quantum or classical reflection. In {\it
quantum reflection} an atom is reflected at the long-range
attractive part of the atom-surface potential \cite{HFriedrich02},
whereas an atom is classically reflected at the turning point of the
potential's repulsive branch. Recently, quantum reflection from
solid surfaces has been observed with ultracold metastable Ne
\cite{Shimizu01} and He atoms \cite{Oberst05a}, with a Bose-Einstein
condensate \cite{Pasquini04}, and with $^3$He \cite{DeKieviet03} and
$^4$He \cite{Zhao08} atom beams of thermal energies. In these
experimental studies classical reflection at the repulsive branch of
the potential was considered to be negligible, either because of
deexcitation of the metastable atoms \cite{Shimizu01,Oberst05a},
inelastic scattering or adsorption \cite{Pasquini04}, or surface
roughness \cite{DeKieviet03}. Quantum reflection was also
theoretically studied, using the long-range Casimir-van der Waals
atom-surface potential, indicating that the reflection probability
is only a function of $k_z$, the component of the incident
wave-vector that is perpendicular to the surface
\cite{HFriedrich02}.

Classical reflection of atom beams from solid surfaces has been
studied intensively for decades, see e.g.\ \cite{Hulpke92}. In most
of those investigations, however, clean crystalline surfaces that
are smooth at the atomic level and that have been kept clean under
ultra-high vacuum conditions have been used. In addition, scattering
of He atoms from disordered surfaces has been used to investigate
local perturbations of the surface including ad-atoms, steps,
clusters, etc. \cite{Gerber87b,Poelsema89,Farias98}. For
microscopically rough surfaces it was generally accepted that atoms
would not be coherently reflected, but would undergo diffuse
scattering. The reflection of atom beams from rough surfaces was
investigated in a few experiments only \cite{Okeefe71,Mason87}. More
recently, microscopic surface roughness has been investigated in the
context of the Casimir force between macroscopic bodies (see e.g.\
Refs.\ \cite{Klimchitskaya99,Neto05,Zwol08}) as well as the
Casimir-Polder interaction between an atom and a rough
\cite{Bezerra2000} or a corrugated surface \cite{Doebrich2008}.

\begin{figure}[b]
\includegraphics[scale=0.32]{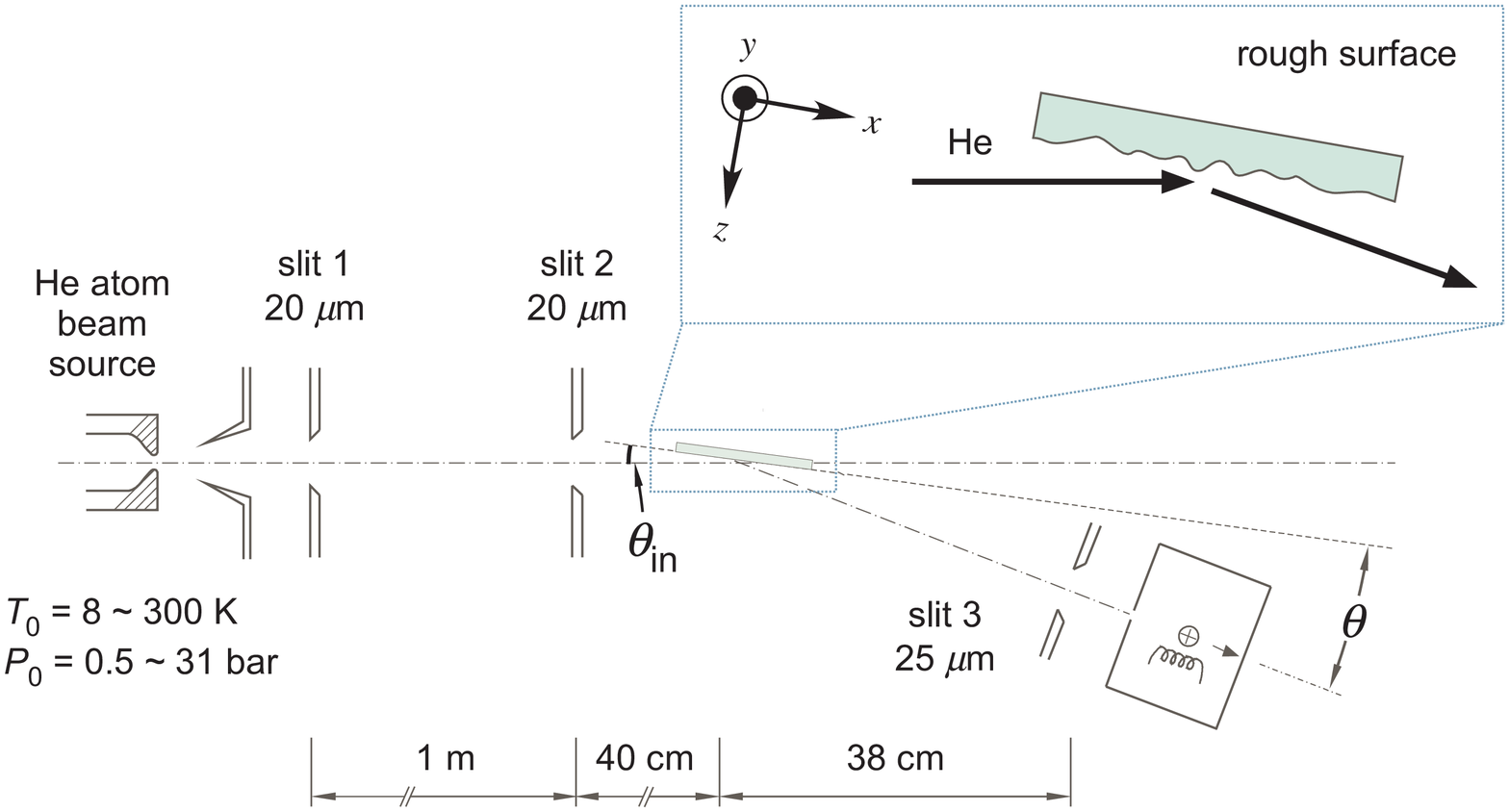}
\caption{(Color online) Scheme of the experimental setup. Angular
patterns are recorded by scanning the detection angle $\theta$ which
is defined with respect to the surface plane. In the inset the
chosen coordinate system is defined; the $xz$-plane and the $z$-axis
are the plane of incidence and the surface normal, respectively,
while the $y$-axis is parallel to the detector pivot axis. }
\label{fig:setup}
\end{figure}

In this article we report on coherent reflection of He atom beams
from rough surfaces. We present reflection probabilities of He atom
beams grazing various planar surfaces: (i) a glass slide for optical
microscopy; (ii) a GaAs wafer; (iii) a chromium surface; and (iv) a
20-$\mu$m-period chromium grating. Even though details of the
reflection probability depend on the material and character of the
surfaces, a general behavior is found for each surface when the
incident wave-vector of the He atom beam is varied. At small $k_z$
the reflection probability is observed to depend only on $k_z$,
whereas at larger $k_z$ the reflection probability also depends on
the wave vector component parallel to the surface; the larger the
parallel wave vector component is, the larger is the reflectivity
for a given value of $k_z$. We attribute the behavior at low $k_z$
to quantum reflection as described recently \cite{Zhao08}, while the
behavior at larger $k_z$ is rationalized in terms of a
classical-reflection model.

The measurements were done with an apparatus described earlier
\cite{Zhao08}. The continuous atom beam is formed in a supersonic
expansion of He gas at stagnation temperature $T_0$ and pressure
$P_0$ through a 5-$\mu$m-diameter orifice into high vacuum. After
passing a skimmer of 500 $\mu$m diameter, the beam is collimated by
two 20-$\mu$m-wide slits (S1 and S2) separated by 100 cm as
indicated in Fig.~\ref{fig:setup}. In combination with the
25-$\mu$m-wide detector-entrance slit (S3), located 78 cm downstream
from the second slit, the angular width of the atom beam is 130
$\mu$rad FWHM (full width at half maximum). The third slit and the
detector (an electron-impact ionization mass spectrometer) are
mounted on a frame which is rotated precisely as indicated in
Fig.~\ref{fig:setup}. The surface under investigation is positioned
such that the (vertical) detector pivot axis is parallel to the
surface and passes through its center. The grazing incidence angle
$\theta_{\rm in}$ and the detection angle $\theta$ are measured with
respect to the surface plane. The reflected beam profiles are
obtained by rotating the detector, namely varying $\theta$, and
measuring the He signal at each angle.

The glass slide is a simple standard microscope slide (ISO Norm
8037/I). It is made out of soda lime glass, is 1 mm thick, and has a
surface area of $76 \times 26$ mm$^2$. It is mounted such that its
shorter direction is parallel to the pivot axis. The commercial GaAs
wafer is cut along the (100) direction and is 50 mm in diameter. The
surface is presumably contaminated with an oxygen layer and slightly
doped with Boron. The 20-$\mu$m-period chromium grating is the same
one used in a previous diffraction experiment \cite{Zhao08}.
Finally, a flat chromium surface of $100 \times 30$ mm$^2$ area is
used for comparison with the grating surface. Both chromium surfaces
are made from commercially available chromium lithography blanks.

No in-situ surface preparation such as Ar-ion sputtering or high
temperature annealing was applied. As the ambient vacuum is about $5
\times 10^{-7}$ mbar we expect each surface to be covered to some
extent with adsorbate molecules from the background gas. Also, all
surfaces were exposed to air for at least several days before
mounting them in the vacuum chamber. Therefore we expect the
surfaces to be oxidized or oxygen covered. Still, for grazing
incidence of the He atom beam intense specular reflection peaks are
observed for each surface.

\begin{figure}[bth]
\includegraphics[scale=0.4]{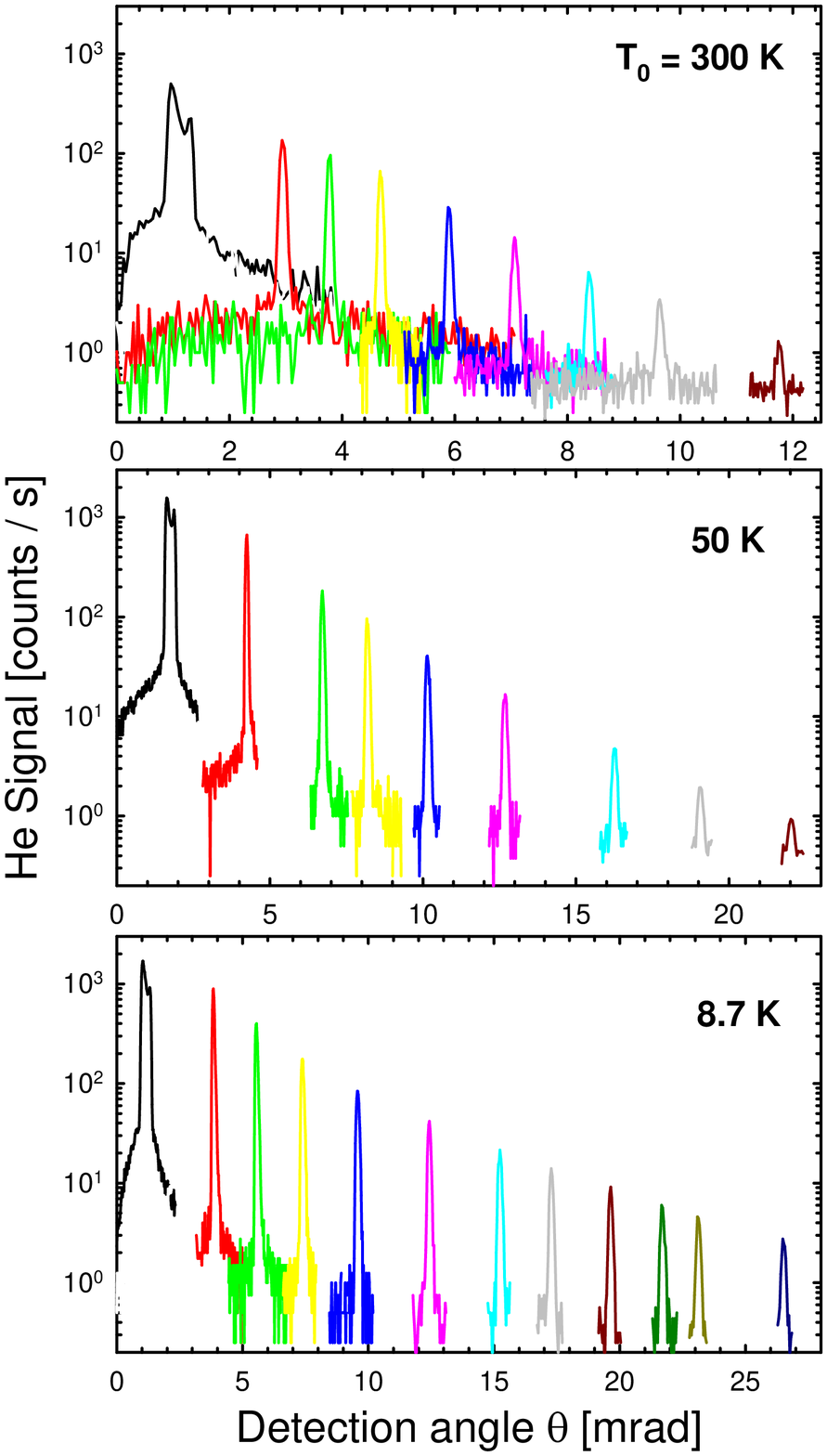}
\caption{(Color online) Angular profiles of He atom beams reflected
from the microscope slide for (a) $T_0 = 300~K $, (b) 50~K, and (c)
8.7~K. In each measurement the incidence angle $\theta_{\rm in}$ is
identical to the detection angle $\theta$ at peak center.}
\label{fig:angprofiles}
\end{figure}

Measurements were made for three stagnation temperatures $T_0 =
300$, 50, and 8.7 K corresponding to incidence wave-vectors $k$ of
112, 46, and 18 nm$^{-1}$, respectively. To maintain a high atom
flux and narrow velocity distribution in the beam and to avoid
cluster formation the stagnation pressure $P_0$ was adjusted to $P_0
= 31$, 26, and 0.5 bar, respectively. Under these conditions the
relative total incident He signals as observed without a surface in
the beam path are 5.0 : 5.5 : 1.0, for $T_0 = 300$, 50, and 8.7 K,
respectively.

\begin{figure}[bth]
\includegraphics[scale=0.65]{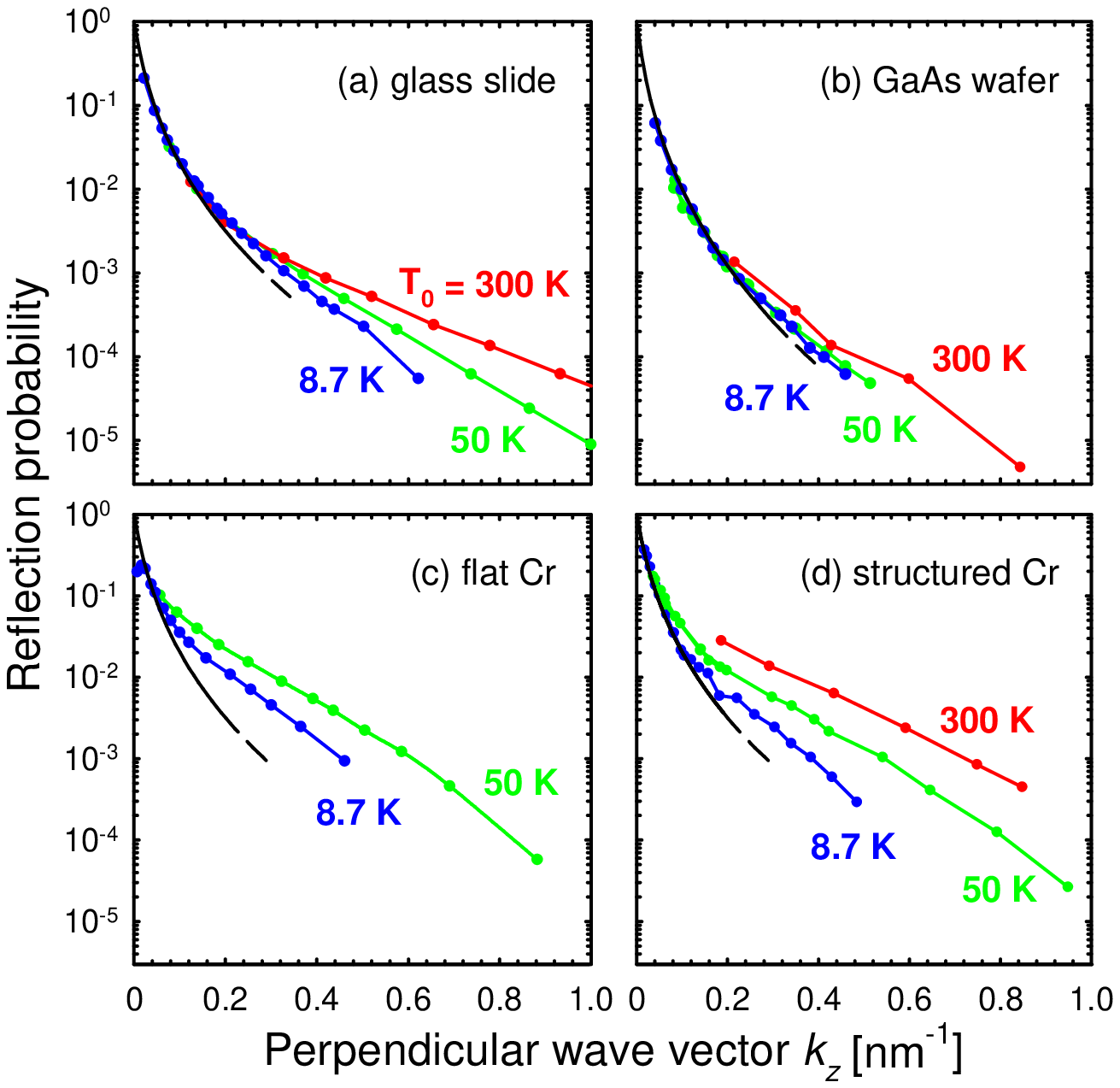}
\caption{(Color online) Reflection probabilities for He atom beams
at source temperatures of 300, 50, and 8 K for different surfaces:
(a) glass slide, (b) GaAs wafer, (c) flat Cr, and (d)
microstructured Cr surface. The black lines are fits by a quantum
reflection calculation. The colored lines connecting the data points
just serve as a guide to the eye.} \label{fig:reflectivity}
\end{figure}

Fig.~\ref{fig:angprofiles} shows angular profiles of the He atom
beam reflected from the microscopy slide at various incidence angles
for the three stagnation conditions. In each series the reflected
peak height decreases by orders of magnitude as the incidence angle
is increased. In addition, broad pedestals that get larger as the
incidence angle decreases are observed under the narrow peaks. We
attribute the broad pedestals to incoherent diffuse scattering in
contrast to the main peaks which are reflected coherently as
evidenced by the observation of diffraction patterns \cite{Zhao08}.
A double peak structure along with a significant broadening of the
main peaks appears for $\theta_{\rm in} \leq 1$ mrad. We tentatively
attribute the former to near-field diffraction at the second slit,
while the broadening is due to a slight curvature of the glass
slide.

The reflection probabilities are determined from the integrated
intensity of the reflected peak normalized to the peak area of the
incident beam. To determine the reflection probability of the
grating surface the sum of all diffraction-peak areas is normalized
to the peak area of the incident beam and multiplied by two, thereby
accounting for the 50\% chromium coverage of the grating surface
\cite{Zhao08}. To allow for comparison between different source
conditions the reflection probabilities are plotted in
Fig.~\ref{fig:reflectivity} as a function of $k_z = k \sin
\theta_{\rm in}$. For the glass slide
(Fig.~\ref{fig:reflectivity}(a)), when $k_z$ is smaller than about
0.3 nm$^{-1}$, the reflection probability is a function of $k_z$
only, and independent of the magnitude of the wave-vector $k$. In
this small-$k_z$ regime the reflection probability decreases steeply
from 22\% at $k_z$ = 0.02 nm$^{-1}$ to about 0.2\% at $k_z$ = 0.3
nm$^{-1}$. For $k_z$ larger than 0.3 nm$^{-1}$, the reflection
probability curves for different stagnation temperatures $T_0$,
i.e., different incidence wave-vectors $k$, start to fan out. In
this regime, for a given $k_z$, the observed reflection probability
appears to increase with increasing parallel wave-vector component
$k_x$.

The steep decrease at small $k_z$ is explained well by quantum
reflection at the attractive branch of the atom-surface interaction
potential \cite{Zhao08}, known as the Casimir-van der Waals
potential, approximated by $V(z) = -C_3 l / [(z+l)z^3]$
\cite{HFriedrich02}. Here, $C_3$ is the van der Waals coefficient,
$z$ denotes the distance between the atom and the surface, and $l$
is a characteristic length that is proportional to the transition
wavelength between the electronic ground state and the first excited
state of the atom ($l = 9.3$ nm for He). The black lines in
Fig.~\ref{fig:reflectivity} are quantum reflection probabilities
obtained by numerically solving the 1-dimensional Schr\"odinger
equation for the attractive potential $V(z)$ with $C_3$ being the
only fit parameter. For the glass slide the best fit to the steep
decrease at small $k_z$ is found for a $C_3$ value of $3 \times
10^{-50}$ Jm$^3$.

The reflection probabilities of the GaAs wafer, the flat chromium
surface and the periodic chromium surface are plotted in
Fig.~\ref{fig:reflectivity}(b)-(d). The black lines in
Fig.~\ref{fig:reflectivity}(b)-(d) represent quantum reflection
calculations with $C_3$ = 5, 3, and 3 $\times 10^{-50}$ Jm$^3$,
respectively. The observed reflection probabilities agree well with
the quantum reflection model until $k_z \simeq$ 0.2 nm$^{-1}$ for
the GaAs wafer and $k_z \simeq$ 0.05 nm$^{-1}$ for the chromium
surfaces. Beyond these values, the observed reflection probabilities
start to deviate from the quantum reflection probabilities and to
spread out for the different stagnation temperatures. The degree of
this fanning out varies for the different surfaces: It is smallest
for the GaAs wafer; larger for the glass slide; and the largest for
both chromium surfaces. It is noteworthy that this trend coincides
with the hierarchy of surface roughness determined independently by
qualitative AFM measurements. These measurements indicate the
largest root-mean-square surface roughness for the chromium surfaces
and the smallest one for GaAs with the glass slide in between.

The combination of a single parameter dependence at small $k_z$ and
a fanning out at larger $k_z$ is a general feature for all surfaces
we have used in reflectivity measurements. To explain this
observation we attribute the behavior at larger $k_z$ to reflection
from the repulsive branch of the atom-surface potential, which we
refer to as classical reflection to emphasize the contrast to
quantum reflection at small $k_z$.

\begin{figure}[tbh]
\includegraphics[scale=0.55]{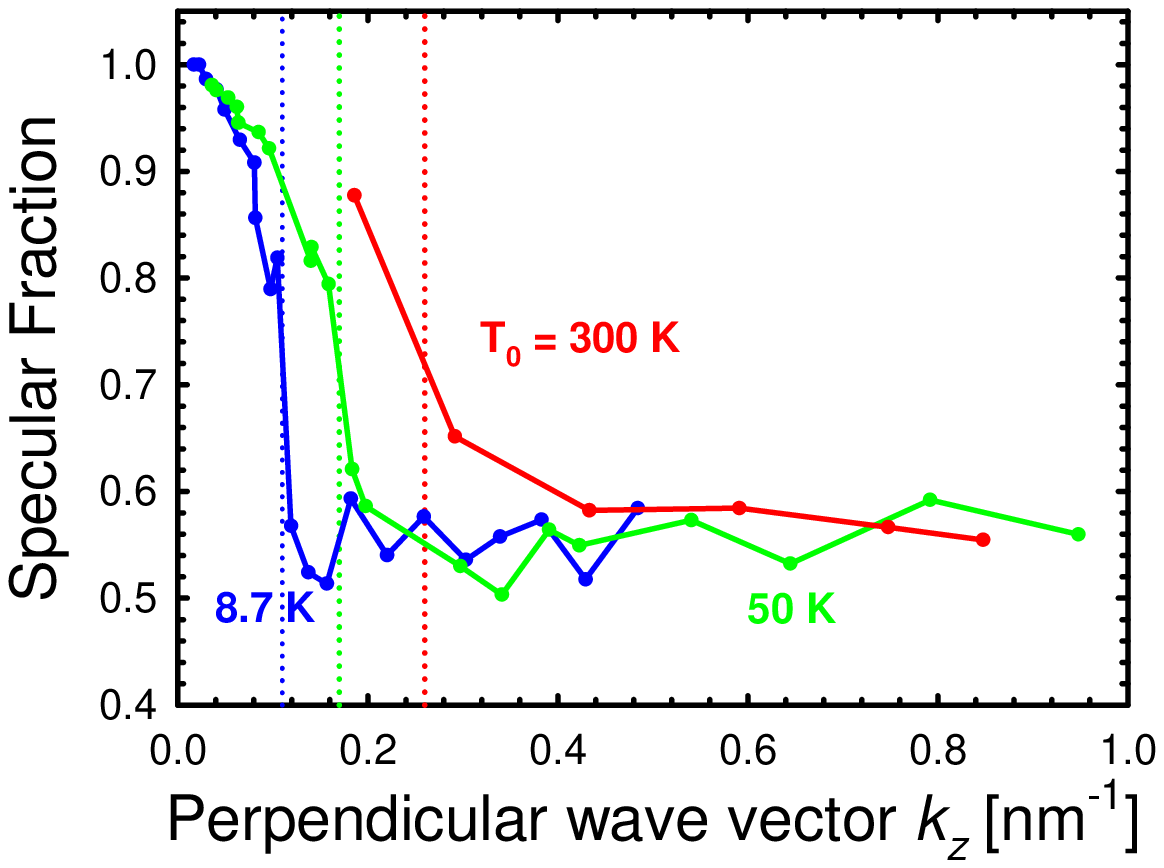}
\caption{(Color online) Ratio of specular peak intensity to the sum
of all diffraction peak intensities observed with the
microstructured Cr surface. The dashed vertical lines mark the
critical values $k^\ast_z$ at which the negative-first-order
diffraction peaks appear.} \label{fig:SpeculPopulation}
\end{figure}

In the following a qualitative explanation of the increase of
classical reflectivity with increasing $k_x$ for a given $k_z$ is
described. Atom scattering from a rough repulsive surface potential
can be understood as averaged diffraction patterns from the
multitude of spatial frequencies within the Fourier spectrum of the
rough surface \cite{Henkel97}. The (non-specular) diffraction peaks
are averaged out leading to a diffusive background signal that does
not contribute to the total reflectivity as this is determined from
the specular peak intensity only. Therefore, the larger the specular
fraction (defined as the ratio of the specular peak intensity to the
sum of all peak intensities) is, the higher is the reflectivity. To
get an idea of how the specular fraction depends on $k_z$ and $k_x$
for He atoms scattering from one of our surfaces, we have analyzed
the diffraction patterns observed with the chromium grating used in
Fig.~\ref{fig:reflectivity}(d) \cite{Zhao08}. In
Fig.~\ref{fig:SpeculPopulation} the specular fraction is plotted
against $k_z$ for the three source conditions.

When $k_z$ is large, the specular fraction stays between 0.5 and 0.6
and is similar for the different $k_x \simeq k$. As $k_z$ decreases
for a given $k_x$, the specular fraction starts to increase at a
certain threshold value and approaches unity at $k_z = 0$ nm$^{-1}$.
The vertical lines in Fig.~\ref{fig:SpeculPopulation} mark the
critical values $k^\ast_z=\sqrt{4\pi k /d}$ at which the negative
first order peak disappears for a given $k$ \cite{Zhao08}, where $d$
is the grating period. Apparently, the observed increase of the
specular fraction coincides with the disappearance of the negative
first order peak. For $k_z$ smaller than the largest critical value
we find a regime where, for a given $k_z$, the specular fraction
increases with $k_x$. This dependence of the specular fraction on
$k_x$ ends at the smallest $k_z$ (less than about $k_z \simeq$ 0.05
nm$^{-1}$), where the curves converge again approaching unity.

The specular fraction approaching unity corresponds to a suppression
of the diffraction peaks. This suppression was discussed previously
for diffraction of atoms from a soft corrugated potential at grazing
incidence \cite{Henkel99}. The physical picture is that at near
grazing incidence many periods of a soft potential are probed by the
atom during its bounce from the surface. This results in an
averaging out of the pase shifts along the various paths thereby
effectively suppressing the diffraction effect. This phase averaging
effect is expected to increase with decreasing angle and, hence,
with increasing $k_x$ for given $k_z$ just as it is found in the
data shown in Fig.~\ref{fig:SpeculPopulation}.

Within this picture it is conceivable that, qualitatively, the same
specular fraction behavior, exemplified here by a 20 $\mu$m periodic
length, would be found for any single spatial-frequency component of
the rough surface spectrum. The relevant $k_z$ scale, however, will
vary because the critical values $k^\ast_z$ depends on the periodic
length $d$; for a short periodic length component, the behavior
described above occurs at large $k_z$, while it happens at small
$k_z$ for a long periodic length component. Hence, averaging the
specular fraction over a range of periodic lengths for a given $k_z$
is qualitatively equivalent to averaging the curves of
Fig.~\ref{fig:SpeculPopulation} over a range of $k_z$. This results
in a larger specular fraction and, hence, larger reflectivity of a
rough surface with increasing $k_x$ at given $k_z$.

In summary, we observed coherent reflection of thermal He atom beams
from microscopically rough surfaces of glass, GaAs, and Cr. For
small $k_z$ the reflection probability is found to be a universal
function of $k_z$ that is modeled well by quantum reflection
\cite{Zhao08}. For larger $k_z$ the reflection probability is found
to increase with $k_x$ for a given $k_z$. The latter behavior has
been discussed qualitatively in terms of an effective averaging-out
of the surface roughness. For a quantitative analysis an improved
theoretical model will be needed accounting for the actual potential
between an atom and a rough surface which could be obtained by
extending the theory for the periodically corrugated surface
\cite{Kirsten89,Dalvit2008} to the randomly rough surface.

B.S.Z.\ acknowledges support by the Alexander von Humboldt
Foundation and by the Korea Research Foundation Grant funded by the
Korean Government (KRF-2005-214-C00188). We thank Markus Heyde for
help with the AFM measurements, Stephan Schulz for supplying the
chromium surfaces, and Samuel A.\ Meek for support with the computer
code.

\end{document}